\documentclass[a4paper]{jpconf}
\usepackage{amssymb}
\usepackage{graphicx}

\newcommand{\be}{\begin{equation}}
\newcommand{\ee}{\end{equation}}

\newcommand{\bea}{\begin{eqnarray}}
\newcommand{\eea}{\end{eqnarray}}

\begin{document}
\title{GPU Computing in Bayesian Inference of Realized Stochastic Volatility Model}

\author{Tetsuya Takaishi}

\address{Hiroshima University of Economics, Hiroshima 731-0192, JAPAN}

\ead{tt-taka@hue.ac.jp}

\begin{abstract}
The realized stochastic volatility (RSV) model that utilizes the realized volatility  as additional information
has been proposed to infer volatility of financial time series. 
We consider the Bayesian inference of the RSV model by the Hybrid Monte Carlo (HMC) algorithm. 
The HMC algorithm can be parallelized and thus performed on the GPU for speedup. 
The GPU code is developed with CUDA Fortran. We compare the computational time in performing the HMC algorithm on GPU (GTX 760) and CPU (Intel i7-4770 3.4GHz) 
and find that the GPU can be up to 17 times faster than the CPU. 
We also code the program with OpenACC and find that appropriate coding can achieve the similar speedup with CUDA Fortran. 
\end{abstract}

\vspace{-5mm}
\section{Introduction}
Since volatility of asset returns plays an important role to manage financial risk, 
measuring values of volatility is an important task in empirical finance.
Since one can not directly observe volatility in the financial markets, 
one needs to use a certain estimation technique such as volatility modeling.
The first promising volatility modeling was introduced in 1982 by Engle\cite{ARCH}. 
The model introduced by him is called the AutoRegressive Conditional Heteroscedasticity (ARCH) model.
Soon the ARCH model was extended by Bollerslev to the generalized ARCH (GARCH) model\cite{GARCH}.
An alternative to the GARCH model is the stochastic volatility (SV) model\cite{Taylor} which allows the volatility to be
a stochastic process.
Recently an extended SV model which utilizes the realized volatility (RV)\cite{RV1,RV2}
constructed by a sum of finely sampled intraday returns
was proposed. This SV-type model is called the realized SV (RSV) model\cite{RSV}.
An advantage of the RSV model over the conventional SV model is that it uses the RV as additional information 
and thus could estimate the daily volatility  more accurately.
Since it is difficult to evaluate the likelihood function of the SV models 
usually the maximum likelihood method is not convenient for the parameter estimation of the SV model. 
The standard estimation algorithm for the SV model is 
the Markov Chain Monte Carlo (MCMC) method based on the Bayesian approach.
Various MCMC algorithms for the SV model have been proposed and tested\cite{SV1}-\cite{Takaishi3}. 

In this study we perform the MCMC method on graphics processing unit (GPU).
GPU computing is able to perform computations in parallel that results in speeding up the computational performance and 
applied for various scientific computations, e.g. Ising model simulations\cite{Ising}-\cite{Ising4}.
Since in general MCMC algorithms are performed sequentially,  
not all MCMC algorithms can be easily parallelized. 
We use the hybrid Monte Carlo (HMC) algorithm\cite{HMC} that can be implemented in parallel.
We perform the HMC algorithm on GPU and CPU for the parameter estimations of the RSV model and compare thier computational performance.
The GPU we used is NVIDIA's GeForce GTX 760 and coding for GPU was done using the PGI CUDA Fortran\cite{PGI}. 
We also code the program using OpenACC\cite{OpenACC} which enables us to do directive based programing and
investigate performance results.

\section{Realized Stochastic Volatility Model}

The realized stochastic volatility (RSV) model introduced by Takahashi {\it et al.}\cite{RSV} is  written as
\bea 
y_t & = &\exp(h_t/2)\epsilon_t, \hspace{5mm} \epsilon_t \sim N(0,1), \\
\ln RV_t & = & \xi + h_t + u_t, \hspace{5mm} u_t \sim N(0,\sigma_u^2), \\
h_{t+1}& = & \mu +\phi(h_t-\mu)+\eta_t,\hspace{5mm}  \eta_t \sim N(0,\sigma_{\eta}^2), 
\eea
where $y_t$ and $RV_t$ for $t=1,...,T$ are  a daily return and a daily realized volatility RV at time $t$ respectively, 
and $h_t$ is a latent volatility defined by $\ln \sigma_t^2$.
The model parameters that we have to estimate are ${\bf \theta }=\phi,\mu,\xi,\sigma_{\eta}^2,\sigma_u^2$.
We apply the Bayesian inference for parameter estimations
and perform the Bayesian inference by the MCMC method. 
The most time consuming part of the MCMC approach for the SV-type models is  
volatility update\cite{SV1}. 
In order to improve the volatility update  several MCMC approaches have been developed, 
e.g. multi-move sampler\cite{SV2,Multimove} and HMC algorithm\cite{Takaishi1,Takaishi2,Takaishi3}.
In this study we use the HMC algorithm for the volatility update of the RSV model. 

\section{Hybrid Monte Carlo Algorithm}
The HMC algorithm\cite{HMC} appeared for the large scale MCMC simulations of
the lattice Quantum Chromo Dynamics (QCD) calculations\cite{LatticeBook} for the first time
and has been the standard MCMC algorithm of the lattice QCD calculations.
The HMC algorithm combines the molecular dynamics (MD) simulation and the Metropolis accept/reject test. 
Since the HMC algorithm is a global algorithm that variables we consider are updated simultaneously.
This means that the variables can be updated in parallel.
For the RSV model we update volatility variables by the HMC algorithm.
The basic HMC algorithm is as follows.
First, candidates for next volatility variables in Markov chain 
are obtained by solving 
the Hamilton's equations of motion in fictitious time $\tau$,
\bea
\label{eq:H1}
\frac{dh_i}{d\tau}& =& \frac{\partial H}{\partial p_i},  \\
\frac{dp_i}{d\tau}& =& -\frac{\partial H}{\partial h_i},
\label{eq:H2}
\eea
where $p_i$ for $i=1,...,T$ is a conjugate momentum to $h_i$.
The Hamiltonian $H$ is defined by 
$H(p,h)=\frac12 \sum_i^n p_i^2 - \ln f(h,\theta)$,
where $f(h,\theta)$ is the conditional posterior density of the RSV model\cite{RSV}.
Since in general eqs.(\ref{eq:H1})-(\ref{eq:H2})
can not be solved analytically, they are  integrated numerically through the MD simulation.
The conventional integrator for the MD simulation in the HMC  algorithm
is the 2nd order leapfrog integrator\cite{HMC} given by
\bea
\label{eq:2LF1}
& h_i(\tau +\delta \tau/2) =  h_i(\tau)+\frac{\delta \tau}{2}p_i(\tau), \\ 
& p_i(\tau+\delta \tau)  =  p_i(\tau)- \delta \tau \frac{\partial H}{\partial h_i}, \\ 
\label{eq:2LF2}
& h_i(\tau +\delta \tau)=  h_i(\tau+\delta \tau/2)+\frac{\delta \tau}{2}p_i(\tau+\delta \tau),
\label{eq:2LF3}
\eea
where $i=1,...,T$ and $\delta \tau$ stands for the step size. 
The higher order or other integrators\cite{Forest}-\cite{HOHMC3} can be also used in the HMC algorithm if necessary.
Eqs.(\ref{eq:2LF1})-(\ref{eq:2LF3}) are repeatedly performed $k$ times and then the total integration length $l$
becomes $l=k\times \delta \tau$.
After the MD simulations we obtain new volatility and conjugate momentum variables, denoted as
$h_i^\prime= h(\tau+l)$ and $p_i^\prime=p(\tau+l)$.
The new volatility variables $h_i^\prime$ are accepted at the Metropolis step with
a probability $\sim \min\{1,\exp(-\Delta H)\}$ where
$\Delta H= H(p^{\prime},h^{\prime})-H(p,h)$.

\section{GPU Coding Environment}
We used the NVIDIA GeForce GTX760 for GPU computing. 
Table 1 shows the specifications of the GTX760\cite{GTX760}.
\begin{table}[t]
\vspace{-2mm}
\centering
\caption{GTX 760 Specifications\cite{GTX760}}
\begin{tabular}{l|r}
\hline
\multicolumn{2}{c}{GPU Engine Specs} \\
\hline
CUDA Cores & 1152 \\
Base Clock (MHz) & 980 \\
Boost Clock (MHz) & 1033 \\
\hline
\multicolumn{2}{c}{Memory Specs} \\
\hline
Memory Speed & 6.0 Gpbs \\
Memory Config & 2048MB \\
Memory Interface & GDDR5 \\
Memory Interface Width & 265-bit \\
Memory Bandwidth (GB/sec) & 192.2 \\
\hline
\end{tabular}
\vspace{-2mm}
\end{table}
The original HMC code of the RSV model for a single CPU was developed in \cite{Takaishi4}.
Our codes for GPU were developed using the PGI CUDA Fortran\cite{PGI} with CUDA 6.0 drivers.
The PGI CUDA Fortran also supports OpenACC\cite{OpenACC}.
All programs are developed with single precision.
We also execute a code on CPU (Intel i7-4770 3.4GHz) to compare performance between GPU and CPU.

\section{HMC Algorithm by CUDA Fortran}

To perform the HMC algorithm on GPU we make a code by CUDA Fortran.
Each equation of eqs.(\ref{eq:2LF1})-(\ref{eq:2LF3}) is assigned to a kernel executed on the GPU as follows.
\bea
\label{eq:kernel1}
&  h_i(\tau +\delta \tau/2) =  h_i(\tau)+\frac{\delta \tau}{2}p_i(\tau),  \hspace{5mm} & \Leftarrow \mbox{\sf Kernel 1} \\ 
&  p_i(\tau+\delta \tau)  =  p_i(\tau)- \delta \tau \frac{\partial H}{\partial h_i} & \Leftarrow \mbox{\sf Kernel 2} \\ 
\label{eq:kernel2}
&  h_i(\tau +\delta \tau) =  h_i(\tau+\delta \tau/2)+\frac{\delta \tau}{2}p_i(\tau+\delta \tau) & \Leftarrow \mbox{\sf Kernel 3} 
\label{eq:kernel3}
\eea
For instance eq.(\ref{eq:kernel1}) which integrates $h_i$ for $i=1,...,T$, denoted by "Kernel 1" 
is coded by CUDA Fortran and executed on the GPU in parallel for $i=1,...,T$. 
After the execution of the Kernel 1, Kernels 2 and 3 are also executed. 
The Kernels 1-3 form an elementary step of the MD simulation.
In this study we executed the elementary step 10000 times and measure an average execution time of the elementary step.
In order to make a comparison between GPU and CPU we also measure an average execution time of the elementary step on CPU.

\begin{figure}[t]
\begin{minipage}{0.5\hsize}
\begin{center}
\includegraphics[height=5cm]{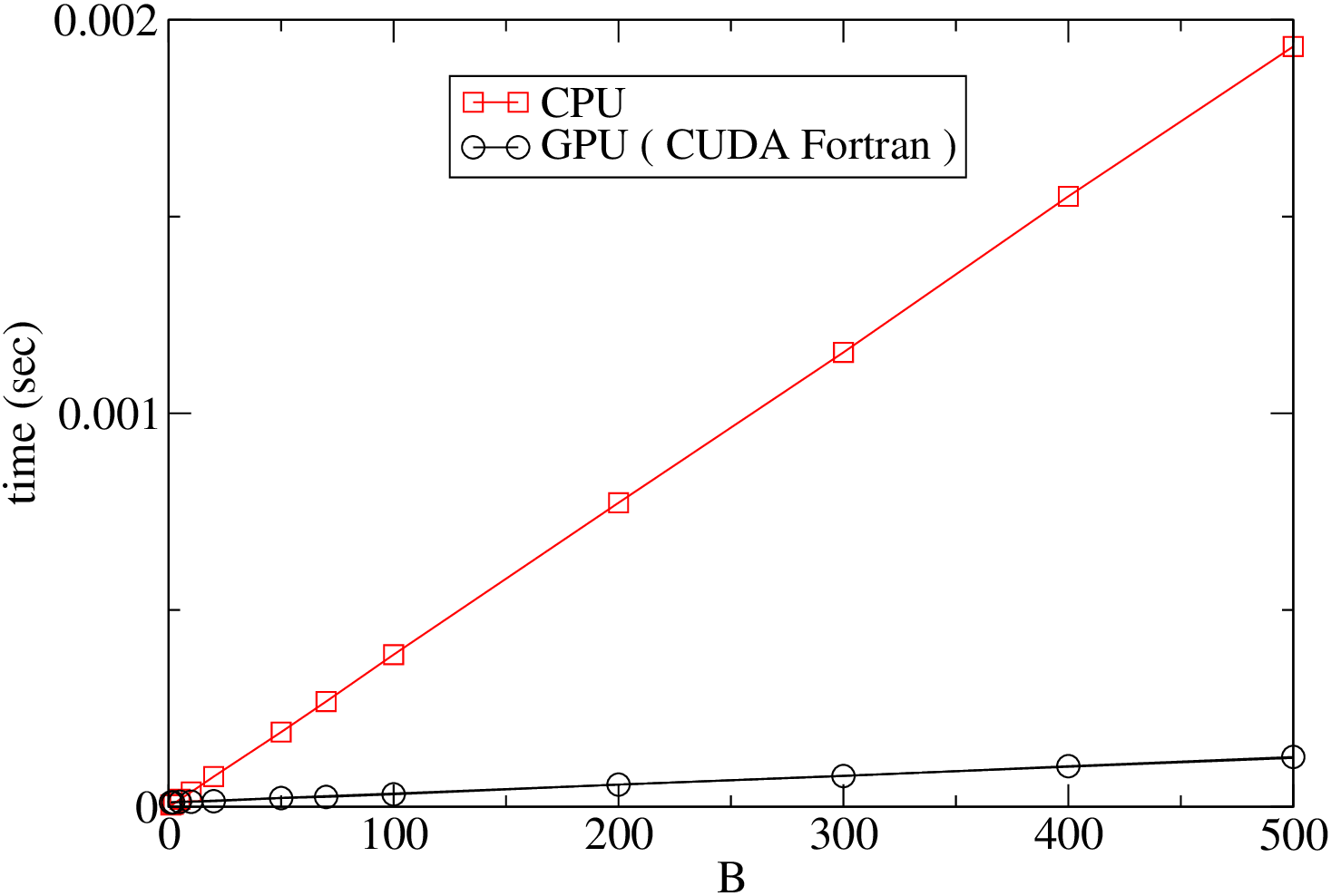}
\caption{ Average execution time of the elementary step on GPU and CPU  as a function of $B$.
}
\label{fig:dH}
\end{center}
\end{minipage}
\hspace{3mm}
\begin{minipage}{0.5\hsize}
\begin{center}
\centering
\includegraphics[height=5cm,keepaspectratio=true]{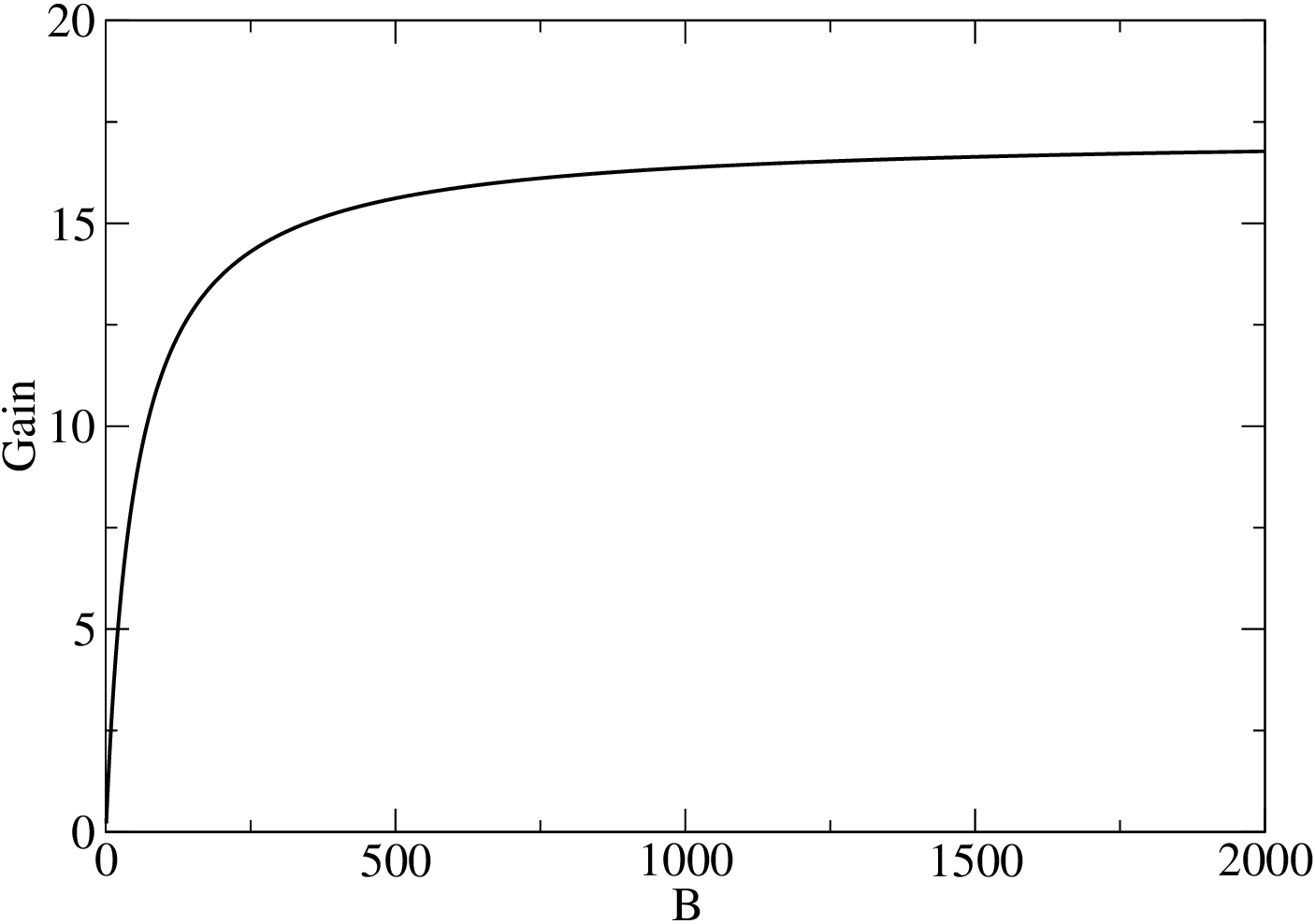}
\caption{Gain defined by $f_{CPU}(B)/f_{GPU}(B)$ as a function of $B$.
}
\label{fig:Acc}
\end{center}
\end{minipage}
\end{figure}

Fig.1 shows the average execution time of the elementary step as a function of the number of volatility variables or the size of time series where 
$B$ has a relationship as $T= 512\times B$. In the GPU computing we set the thread size to 512.
From Fig.1 we recognize that the average execution time increases almost linearly with B for both GPU and CPU.
We fit the results with a linear function of $f(B)=A + C \times B$ for both GPU and CPU. 
The fitting results are summarized in Table 2.
Here we define the speedup of GPU over the CPU by $Gain=f_{CPU}(B)/f_{GPU}(B)$.
Fig.2 shows the Gain as a function of $B$. 
The Gain increases with $B$ and goes to $C_{CPU}/C_{GPU}=17.2$ in the limit of $B \rightarrow \infty$.

\begin{table}[t]
\vspace{-2mm}
\centering
\caption{Fitting parameters}
\begin{tabular}{l|c|c}
\hline
 & A & C \\
\hline
GPU (CUDA Fortran)  & $1.13\times 10^{-5}$ & $2.25\times 10^{-7}$ \\
CPU (Intel i7-4770 3.4GHz) & $-1.42\times 10^{-6}$ & $3.87\times 10^{-6}$\\
\hline
\end{tabular}
\vspace{-2mm}
\end{table}

\section{HMC algorithm by OpenACC}
OpenACC enables us to do directive based programming for GPU that can greatly reduce coding effort.
We insert OpenACC directives into the existing Fortran program and actually in this study
we used the program developed for the RSV model in \cite{Takaishi4}.
The following is the schematic diagram for the OpenACC coding. 

\vspace{2mm}
{\bf 
!\$acc data copy(h,p) 

\vspace{1mm}
!\$acc kernels
\bea
\label{eq:Open1}
& h_i(\tau +\delta \tau/2) =  h_i(\tau)+\frac{\delta \tau}{2}p_i(\tau)   \hspace{6cm} \\ 
& p_i(\tau+\delta \tau)   =  p_i(\tau)- \delta \tau \frac{\partial H}{\partial h_i}  \hspace{6.5cm}  \\ 
\label{eq:Open2}
& h_i(\tau +\delta \tau)  =  h_i(\tau+\delta \tau/2)+\frac{\delta \tau}{2}p_i(\tau+\delta \tau)  \hspace{4.5cm} 
\label{eq:Open3}
\eea

!\$acc end kernels

\vspace{1mm}
!\$acc end data
\vspace{2mm}
}

Eqs.(\ref{eq:Open1})-(\ref{eq:Open3}) between "!\$acc kernels" and "!\$acc end kernels"
are automatically translated to a GPU code and performed on GPU.
The data directive "!\$acc data copy(h,p)" specifies the variables ( here $h$ and $p$ ) that are used in the GPU code 
between "!\$acc data copy(h,p)" and "!\$acc end data". This avoids unnecessary data transfer between CPU and GPU.
Actually in order to measure an average execution time as in the previous section we repeat 10000 times 
the code from "!\$acc kernels" to "!\$acc end kernels" and if no data directive is inserted to the program
we can not achieve the speedup as the code by CUDA Fortran.
Fig.3  shows the average execution time as a function of $B$.
The squares (circles) indicate the average execution time of the OpenACC code with (without) the data directive.
We find that the average execution time of the OpenACC code with the data directive is similar with 
that of the CUDA Fortran. 
On the other hand, without the data directive the average execution time takes more time than that with the data directive. 

\begin{figure}[t]
\vspace{-2mm}
\begin{center}
\includegraphics[height=5.5cm]{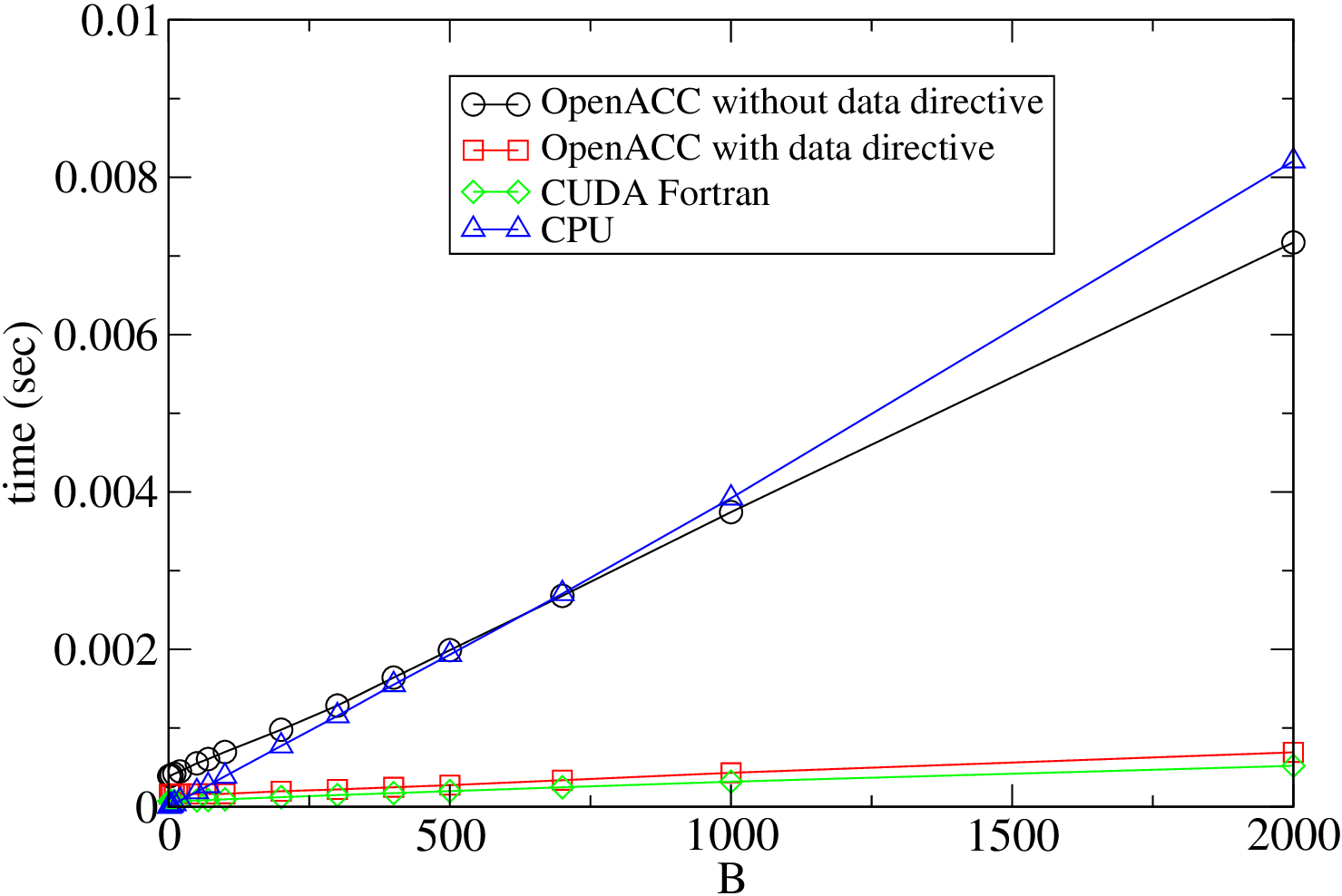}
\end{center}
\vspace{-3mm}
\caption{ 
Average execution time of the elementary step of GPU computing by OpenACC with (without) data directive. 
}
\end{figure}

\section{Conclusion}
We have performed the HMC algorithm in the Bayesian estimation of the RSV model 
on GPU (GTX 760) using CUDA Fortran.
It is found that the GPU can be up to 17 times faster than the CPU (Intel i7-4770 3.4GHz) when the size of time series is big.
We have also coded an HMC program for GPU with the OpenACC that enables us to do directive based programming 
for GPU computing and found that the OpenACC program with appropriate coding can achieve the similar speedup with CUDA Fortran. 

\section*{Acknowledgement}
Numerical calculations in this work were carried out at the
Yukawa Institute Computer Facility
and at the facilities of the Institute of Statistical Mathematics.
This work was supported by JSPS KAKENHI Grant Number 25330047.

\end{document}